\documentclass[10pt,conference]{IEEEtran}

\usepackage{color}

\ifx\pdfoutput\undefined
\usepackage{graphicx}
\graphicspath{{eps_figs/}}
\else
\usepackage[pdftex]{graphicx}
\graphicspath{{pdf_figs/}}
\fi

\usepackage{latexsym,amssymb,amsmath,ifthen,afterpage,cite}

\newcommand{\Fig}[1]{Fig.~\ref{#1}}

\newcommand{\eqdef}{\stackrel{\scriptscriptstyle\bigtriangleup}{=} }
\newcommand{\R}{\mathbb{R}}

\newcommand{\B}{{\mathcal{B}}}

\newcommand{\calX}{\mathcal{X}}

\newcommand{\calU}{\mathcal{U}}

\newcommand{\X}{\bf{X}}
\newcommand{\x}{{\bf x}}

\newcommand{\E}{\operatorname{E}}

\newcounter{examplecntr}
{\begin{trivlist}\small\item[]\refstepcounter{examplecntr}%
 {\bfseries Example~\theexamplecntr%
  \ifthenelse{\equal{#1}{}}{}{ (#1)}.
}}%
{\end{trivlist}}

\newcounter{theoremcntr}
{\begin{trivlist}\item[]\refstepcounter{theoremcntr}%
{\bfseries Theorem~\thetheoremcntr%
  \ifthenelse{\equal{#1}{}}{}{ (#1)}.
}}%
{\hfill$\Box$\end{trivlist}}

\newcommand{\pos}[2]{\makebox(0,0)[#1]{#2}}

\let\oldbibliography\thebibliography
\renewcommand{\thebibliography}[1]{%
  \oldbibliography{#1}%
  \setlength{\itemsep}{0.68pt}%
}

\IEEEoverridecommandlockouts
 
\begin{document}
\DeclareGraphicsExtensions{.pdf}

\title{An Importance Sampling Scheme\\ for Models in a Strong External Field
}

%\title{Monte Carlo Methods in the\\ Dual Factor Graph of the Ising Model
%} 
%AN IMPORTANCE SAMPLING SCHEME FOR MODELS IN A STRONG %EXTERNAL FIELD
\author{Mehdi Molkaraie \\
Universitat Pompeu Fabra \\
08018 Barcelona, Spain \\
\tt{mehdi.molkaraie@alumni.ethz.ch}
%\thanks{This work has been funded in part by the
%European Research Council under ERC grant agreement
%259663, and by the Spanish Ministry of Economy and Competitiveness under
%grant TEC2012-38800-C03-03.}
}

\maketitle 
 
\begin{abstract} 
We propose Monte Carlo methods to estimate the 
partition function of the two-dimensional 
Ising model in the presence of an external 
magnetic field. 
The estimation is done  
in the dual of the Forney factor graph representing the model. The proposed methods can efficiently  
compute an estimate of the partition function in a wide range of
model parameters. As an example, we consider models that are in a strong external 
field.
\end{abstract} 
 
\section{Introduction} 
 
In~\cite{MoLo:ISIT2013}, the authors showed that for 
the two-dimensional (2D) Ising
model, at low temperature,  
Monte Carlo methods converge faster in the dual Forney factor graph than in the original (primal) factor graph.
Monte Carlo methods based on the dual factor graph were also
proposed in~\cite{MoLo:ISIT2013}
to estimate of the partition function of the 2D Ising model 
%(with constant or with spatially varying couplings) 
in the absence of an external magnetic field (see also~\cite{AY:2014}).

In thermodynamic limits, the exact value of the partition function of 
the 2D Ising model, with constant couplings and
in the absence of an external field, was first calculated by 
Onsager~\cite{Onsager:44},~\cite[Chapter 7]{Baxter07}.
However, the 2D Ising model in an arbitrary non-zero external 
field and the three-dimensional (3D) Ising model have remained 
unsolved~\cite{Welsh:90}.

In general, the partition function of 2D models
with arbitrary couplings can be estimated 
using Markov chain Monte 
Carlo methods~\cite{Neal:proinf1993r,BH:10,LoMo:IT2013}.
%At low temperatures, however, 
%Monte Carlo methods usually suffer from critical slowing down. 
%It is well known that at a certain 
%critical temperature, the 2D 
%ferromagnetic
%Ising model undergoes a phase transition;
%below this temperature, variables (spins) have long-range dependencies and
%%Markov chain 
%Monte Carlo methods (based on single spin-flips) do 
%not mix rapidly~\cite{BH:10}. 
In this paper, we consider the problem of estimating the partition function 
of the ``finite-size" 2D  ferromagnetic 
Ising model in a consistent external field.
We propose Monte Carlo methods in the dual of the Forney factor 
graph representing the model that can efficiently 
estimate the partition function. As a special case, we consider models that are 
in a strong external magnetic field.
%which can be used to compute
%the partition function of models
%with pairwise interactions. In this paper, we are mainly concerned with computing the 
%partition function
%of finite-size 2D ferromagnetic Ising models
%%models and 2D ferromagnetic 
%and $q$-state Potts models~\cite{Potts:52}, 

%In our numerical experiments,
%we will also consider 3D 
%ferromagnetic Ising models.
%The importance sampling 
%scheme operates on the dual of the Forney factor graphs representing the models. 

%It must be emphasized that, unlike well-known 
%algorithms, e.g., Gibbs sampling~\cite{GG:srgd1984} and 
%the Swendsen-Wang
%algorithm~\cite{SW:87}, the proposed scheme does not suggest 
%a method to draw samples according to the Boltzmann distribution on factor graphs,
%as sampling is done in the dual domain. 

The paper is organized as follows. In Section~\ref{sec:Ising},
we review the Ising model
and graphical model representations in terms of 
Forney factor graphs. 
Dual Forney factor graphs and 
the  factor graph duality theorem
are discussed in Section~\ref{sec:NFGD}.
The proposed Monte Carlo methods are described in Section~\ref{sec:IS}. 
Numerical experiments are reported in Section~\ref{sec:Num}.

\section{The Ising Model in an External Magnetic Field}
\label{sec:Ising}

\begin{figure}[t]
\setlength{\unitlength}{0.88mm}
\centering
\begin{picture}(81,72)(0,0)
\small
\put(0,60){\framebox(4,4){$=$}}
 \put(4,60){\line(4,-3){4}}
 \put(8,54){\framebox(3,3){}}
\put(4,62){\line(1,0){8}}        \put(8,63){\pos{bc}{$X_1$}}
\put(12,60){\framebox(4,4){}}
\put(16,62){\line(1,0){8}}
\put(24,60){\framebox(4,4){$=$}}
 \put(28,60){\line(4,-3){4}}
 \put(32,54){\framebox(3,3){}}
\put(28,62){\line(1,0){8}}       \put(32,63){\pos{bc}{$X_2$}}
\put(36,60){\framebox(4,4){}}
\put(40,62){\line(1,0){8}}
\put(48,60){\framebox(4,4){$=$}}
 \put(52,60){\line(4,-3){4}}
 \put(56,54){\framebox(3,3){}}
\put(52,62){\line(1,0){8}}       
%\put(56,63){\pos{bc}{$X_3$}}
\put(60,60){\framebox(4,4){}}
\put(64,62){\line(1,0){8}}
\put(72,60){\framebox(4,4){$=$}}
 \put(76,60){\line(4,-3){4}}
 \put(80,54){\framebox(3,3){}}
\put(2,54){\line(0,1){6}}
\put(0,50){\framebox(4,4){}}
\put(2,50){\line(0,-1){6}}
\put(26,54){\line(0,1){6}}
\put(24,50){\framebox(4,4){}}
\put(26,50){\line(0,-1){6}}
\put(50,54){\line(0,1){6}}
\put(48,50){\framebox(4,4){}}
\put(50,50){\line(0,-1){6}}
\put(74,54){\line(0,1){6}}
\put(72,50){\framebox(4,4){}}
\put(74,50){\line(0,-1){6}}
\put(0,40){\framebox(4,4){$=$}}
 \put(4,40){\line(4,-3){4}}
 \put(8,34){\framebox(3,3){}}
\put(4,42){\line(1,0){8}}
\put(12,40){\framebox(4,4){}}
\put(16,42){\line(1,0){8}}
\put(24,40){\framebox(4,4){$=$}}
 \put(28,40){\line(4,-3){4}}
 \put(32,34){\framebox(3,3){}}
\put(28,42){\line(1,0){8}}
\put(36,40){\framebox(4,4){}}
\put(40,42){\line(1,0){8}}
\put(48,40){\framebox(4,4){$=$}}
 \put(52,40){\line(4,-3){4}}
 \put(56,34){\framebox(3,3){}}
\put(52,42){\line(1,0){8}}
\put(60,40){\framebox(4,4){}}
\put(64,42){\line(1,0){8}}
\put(72,40){\framebox(4,4){$=$}}
 \put(76,40){\line(4,-3){4}}
 \put(80,34){\framebox(3,3){}}
\put(2,34){\line(0,1){6}}
\put(0,30){\framebox(4,4){}}
\put(2,30){\line(0,-1){6}}
\put(26,34){\line(0,1){6}}
\put(24,30){\framebox(4,4){}}
\put(26,30){\line(0,-1){6}}
\put(50,34){\line(0,1){6}}
\put(48,30){\framebox(4,4){}}
\put(50,30){\line(0,-1){6}}
\put(74,34){\line(0,1){6}}
\put(72,30){\framebox(4,4){}}
\put(74,30){\line(0,-1){6}}
\put(0,20){\framebox(4,4){$=$}}
 \put(4,20){\line(4,-3){4}}
 \put(8,14){\framebox(3,3){}}
\put(4,22){\line(1,0){8}}
\put(12,20){\framebox(4,4){}}
\put(16,22){\line(1,0){8}}
\put(24,20){\framebox(4,4){$=$}}
 \put(28,20){\line(4,-3){4}}
 \put(32,14){\framebox(3,3){}}
\put(28,22){\line(1,0){8}}
\put(36,20){\framebox(4,4){}}
\put(40,22){\line(1,0){8}}
\put(48,20){\framebox(4,4){$=$}}
 \put(52,20){\line(4,-3){4}}
 \put(56,14){\framebox(3,3){}}
\put(52,22){\line(1,0){8}}
\put(60,20){\framebox(4,4){}}
\put(64,22){\line(1,0){8}}
\put(72,20){\framebox(4,4){$=$}}
 \put(76,20){\line(4,-3){4}}
 \put(80,14){\framebox(3,3){}}
\put(2,14){\line(0,1){6}}
\put(0,10){\framebox(4,4){}}
\put(2,10){\line(0,-1){6}}
\put(26,14){\line(0,1){6}}
\put(24,10){\framebox(4,4){}}
\put(26,10){\line(0,-1){6}}
\put(50,14){\line(0,1){6}}
\put(48,10){\framebox(4,4){}}
\put(50,10){\line(0,-1){6}}
\put(74,14){\line(0,1){6}}
\put(72,10){\framebox(4,4){}}
\put(74,10){\line(0,-1){6}}
\put(0,0){\framebox(4,4){$=$}}
 \put(4,0){\line(4,-3){4}}
 \put(8,-6){\framebox(3,3){}}
\put(4,2){\line(1,0){8}}
\put(12,0){\framebox(4,4){}}
\put(16,2){\line(1,0){8}}
\put(24,0){\framebox(4,4){$=$}}
 \put(28,0){\line(4,-3){4}}
\put(32,-6){\framebox(3,3){}}
\put(28,2){\line(1,0){8}}
\put(36,0){\framebox(4,4){}}
\put(40,2){\line(1,0){8}}
\put(48,0){\framebox(4,4){$=$}}
 \put(52,0){\line(4,-3){4}}
 \put(56,-6){\framebox(3,3){}}
\put(52,2){\line(1,0){8}}
\put(60,0){\framebox(4,4){}}
\put(64,2){\line(1,0){8}}
\put(72,0){\framebox(4,4){$=$}}
 \put(76,0){\line(4,-3){4}}
 \put(80,-6){\framebox(3,3){}}
\end{picture}
%
%\vspace{3ex}
\vspace{2.5ex}
\caption{\label{fig:2DGrid}%
Forney factor graph of the 2D Ising model in 
an external field, where unlabeled normal-size 
boxes represent% factors as in
~(\ref{eqn:IsingA}), 
small boxes represent%factors as in
~(\ref{eqn:IsingH}), and 
boxes containing $``="$ 
symbols are equality constraints.
}
\end{figure}
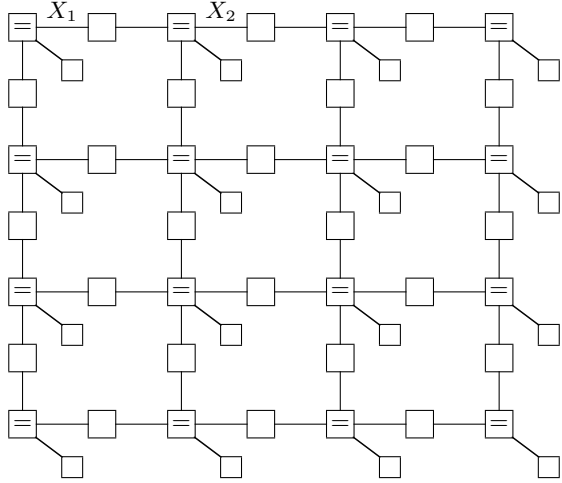

%We consider the problem of computing the partition 
%function of one-dimensional (1D) and finite-size 
%two-dimensional (2D) Ising models.
%In particular, we propose a method
%to compute the partition function of finite-size 2D Ising 
%models at low temperature 
%by performing
%Markov chain Monte Carlo methods on the dual factor 
%graph.
Let $X_1, X_2, \ldots, X_N$ be a collection of discrete random variables
arranged on the sites of a 2D lattice,
as illustrated in Fig.~\ref{fig:2DGrid}, where interactions 
are restricted to 
adjacent (nearest-neighbor) variables. 
Suppose each random
variable takes on values in a finite alphabet $\calX$.
Let $x_i$ represent a
possible realization of $X_i$, $\x$ stand for 
a configuration $(x_1, x_2, \ldots, x_N)$, and $\X$ stand for 
$(X_1, X_2, \ldots, X_N)$. 

In a 2D Ising model, $\calX = \{0,1\}$ and the 
Hamiltonian (the energy function)  of a configuration $\x$ 
is defined as~\cite{Baxter07}
\begin{multline}
\label{eqn:HamiltonianI}
\mathcal{H}(\x) \eqdef -\!\!\sum_{\text{$(k,\ell)\in \B$}}\!\!\!J_{k, \ell}\cdot
\big([x_k = x_{\ell}] - [x_k \ne x_{\ell}]\big)\\ 
- \sum_{m = 1}^N H_m\cdot\big([x_m = 1] - [x_m = 0]\big)
\end{multline}
where $\B$ contains 
all the unordered pairs (bonds) $(k,\ell)$ with non-zero 
interactions and $[\cdot]$ denotes the Iverson 
bracket~\cite[Chapter 2]{GKP:89}, which evaluates to $1$ if the condition in 
the bracket is satisfied and to $0$ otherwise.

The real coupling parameter $J_{k, \ell}$ controls the strength of
the interaction between adjacent variables $(x_k, x_{\ell})$.
The real parameter $H_m$ corresponds to the
presence of an external magnetic field. 
%The model is called ferromagnetic if all coupling parameters 
%are positive, i.e.,  
In this paper, we concentrate on ferromagnetic models,
characterized by $J_{k, \ell} > 0$ for each $(k, \ell) \in \B$.
%In ferromagnetic models, configurations in which adjacent 
%variables with the same value have lower energy levels. 
The external field is assumed to be consistent, i.e., for $1 \le m \le N$, $H_m$ is either assigned 
to all positive or to all negative values.
%If $H_m > 0$, variable $X_m$ tends to have
%value $1$, while $X_m$ tends to have
%value $0$ if $H_m < 0$.

%and controls the strength of
%the interaction between $X_m$ and the field.

%%%%%%%%%%%%%%%%%%%%%%%%%%%%%%%%%%%%%

The probability that the model is in configuration $\x$ is
given by the Boltzmann distribution~\cite{Baxter07}
\begin{equation} 
\label{eqn:Prob}
p_{\text{B}}(\x) = \frac{e^{- \beta\mathcal{H}(\x)}}{Z} 
\end{equation}

Here, the normalization constant $Z$ is the 
\emph{partition function}  $Z = \sum_{\x \in \calX^N} e^{ -\beta\mathcal{H}(\x)}$
and $\beta = 1/k_{\text{B}}T$, where $T$ denotes the temperature 
and $k_{\text{B}}$ is Boltzmann's constant. 

In the rest of this paper, we 
will assume $\beta = 1$. Hence, large values of $J$ and $|H|$ correspond to models at low temperature
and in a strong external field. Boundary conditions
are assumed to be periodic throughout this paper. Thus $|\B| = 2N$.
%defined as $F_{\text{H}} \eqdef -\ln (Z)$.
%%%%%%%%%%%%%%%%%%%%%%%%%%%%%%%%%%%%%

%The Helmholtz free energy is defined as
%\begin{equation} 
%\label{eqn:FreeEnergy}
%F_{\text{H}} \eqdef -\frac{1}{\beta}\ln Z
%\end{equation}

%In the rest of this paper, we will assume $\beta = 1$.
%With this 
%assumptions, e.g., 
%large values of $J$ and $|H|$ correspond to models at low temperature
%and in a strong external %magnetic 
%field. 

For each adjacent pair $(x_k, x_\ell)$, let 
\begin{equation}
\label{eqn:IsingA}
\kappa_{k, \ell}(x_k, x_{\ell}) 
= e^{J_{k, \ell}\cdot\big([x_k = x_{\ell}] - [x_k \ne x_{\ell}]\big)}
\end{equation}
and for each $x_m$
\begin{equation}
\label{eqn:IsingH}
\tau_{m}(x_m) = e^{H_m\cdot\big([x_m = 1] - [x_m = 0]\big)}
\end{equation}

We then define $f: \calX^N \rightarrow \R_{> 0}$ as 
\begin{equation} 
\label{eqn:factorF}
f(\x) \, \eqdef  \!\!\prod_{\text{$(k,\ell)\in \B$}}\!\!\!\kappa_{k, \ell}(x_k, x_{\ell})
 \prod_{m = 1}^N \tau_{m}(x_m)
\end{equation}

The corresponding Forney factor 
graph (normal Factor graph) 
for the factorization
in~(\ref{eqn:factorF}) is shown in~\Fig{fig:2DGrid},
where the boxes labeled ``$=$'' are equality 
constraints~\cite{Forney:01,Lg:ifg2004}. 

From (\ref{eqn:factorF}), the partition function in~(\ref{eqn:Prob})
can be expressed as 
\begin{equation}
\label{eqn:PartFunction}
Z = \sum_{\x \in \calX^N} f(\x) 
\end{equation}

%For small $J$ (i.e., at high temperature) and small $|H|$, 
%the Boltzmann distribution~(\ref{eqn:Prob})
%approaches the uniform distribution. In this case, we can estimate $Z$ efficiently via Monte Carlo methods 
%in the original Forney factor graph in Fig.~\ref{fig:2DGrid}.
To estimate $Z$, we propose
Monte Carlo methods in the dual of the Forney factor graph representing the 
factorization~(\ref{eqn:factorF}). 

%where the only requirement for
%fast convergence is having a strong external field (i.e., large $|H|$).
%In more challenging 
%situations (e.g., models at low temperature or in a strong external field), the Boltzmann distribution
%will be highly non-uniform with isolated modes.  
%The scheme 
%In particular, we apply the proposed method to models in a strong external field .

%\vspace{3.5ex}

%%%%%%%%%%%%%%%%%%%%%%%%%%%%%%%%%

\section{The Dual Model}
\label{sec:NFGD}

We can obtain the dual of~\Fig{fig:2DGrid}, by 
replacing each variable $x$ with its dual
variable $\tilde x$, each 
factor $\kappa_{k, \ell}$ with its 2D discrete Fourier 
transform (DFT), 
%\footnote{Here, $\gamma(\tilde x_1, \tilde x_2)$, the 2D DFT of $\kappa(x_1, x_2)$, 
%is defined as
%\begin{equation*} 
%\gamma(\tilde x_1, \tilde x_2) \eqdef 
%\sum_{x_1\in \calX}\sum_{x_2 \in \calX} \kappa(x_1,x_2)
%e^{-i2\pi(x_1\tilde x_1 + x_2\tilde x_2)/|\calX|}
%\end{equation*}
%where $i $ is the unit imaginary number~\cite{Brace:1999}~\cite[Chapter 10]{Sand:2001}.\vspace{0.25mm}}
each factor $\tau_m$ with its one-dimensional (1D) 
DFT, and each equality 
constraint with an XOR factor~\cite{Forney:01, Forney:11, AY:2011, FV:2011}. 
Note that $\tilde X$ also takes on values in $\calX$.
%Note that, $\tilde X$
%also takes on values in $\calX$.

\begin{figure}[t]
\setlength{\unitlength}{0.88mm}
\centering
\begin{picture}(77,72)(0,0)
\small
\put(0,60){\framebox(4,4){$+$}}
 \put(4,60){\line(4,-3){4}}
 \put(8,54){\framebox(3,3){}}
\put(4,62){\line(1,0){8}}        
\put(12,60){\framebox(4,4){$=$}}
\put(16,62){\line(1,0){8}}
\put(24,60){\framebox(4,4){$+$}}
 \put(28,60){\line(4,-3){4}}
 \put(32,54){\framebox(3,3){}}
\put(28,62){\line(1,0){8}}       
\put(36,60){\framebox(4,4){$=$}}
\put(40,62){\line(1,0){8}}
\put(48,60){\framebox(4,4){$+$}}
 \put(52,60){\line(4,-3){4}}
 \put(56,54){\framebox(3,3){}}
\put(52,62){\line(1,0){8}}       
\put(60,60){\framebox(4,4){$=$}}
\put(64,62){\line(1,0){8}}
\put(72,60){\framebox(4,4){$+$}}
 \put(76,60){\line(4,-3){4}}
 \put(80,54){\framebox(3,3){}}
\put(2,54){\line(0,1){6}}
\put(0,50){\framebox(4,4){$=$}}
\put(2,50){\line(0,-1){6}}
\put(26,54){\line(0,1){6}}
\put(24,50){\framebox(4,4){$=$}}
\put(26,50){\line(0,-1){6}}
\put(50,54){\line(0,1){6}}
\put(48,50){\framebox(4,4){$=$}}
\put(50,50){\line(0,-1){6}}
\put(74,54){\line(0,1){6}}
\put(72,50){\framebox(4,4){$=$}}
\put(74,50){\line(0,-1){6}}
\put(0,40){\framebox(4,4){$+$}}
 \put(4,40){\line(4,-3){4}}
 \put(8,34){\framebox(3,3){}}
\put(4,42){\line(1,0){8}}
\put(12,40){\framebox(4,4){$=$}}
\put(16,42){\line(1,0){8}}
\put(24,40){\framebox(4,4){$+$}}
 \put(28,40){\line(4,-3){4}}
 \put(32,34){\framebox(3,3){}}
\put(28,42){\line(1,0){8}}
\put(36,40){\framebox(4,4){$=$}}
\put(40,42){\line(1,0){8}}
\put(48,40){\framebox(4,4){$+$}}
 \put(52,40){\line(4,-3){4}}
 \put(56,34){\framebox(3,3){}}
\put(52,42){\line(1,0){8}}
\put(60,40){\framebox(4,4){$=$}}
\put(64,42){\line(1,0){8}}
\put(72,40){\framebox(4,4){$+$}}
 \put(76,40){\line(4,-3){4}}
 \put(80,34){\framebox(3,3){}}
\put(2,34){\line(0,1){6}}
\put(0,30){\framebox(4,4){$=$}}
\put(2,30){\line(0,-1){6}}
\put(26,34){\line(0,1){6}}
\put(24,30){\framebox(4,4){$=$}}
\put(26,30){\line(0,-1){6}}
\put(50,34){\line(0,1){6}}
\put(48,30){\framebox(4,4){$=$}}
\put(50,30){\line(0,-1){6}}
\put(74,34){\line(0,1){6}}
\put(72,30){\framebox(4,4){$=$}}
\put(74,30){\line(0,-1){6}}
\put(0,20){\framebox(4,4){$+$}}
 \put(4,20){\line(4,-3){4}}
 \put(8,14){\framebox(3,3){}}
\put(4,22){\line(1,0){8}}
\put(12,20){\framebox(4,4){$=$}}
\put(16,22){\line(1,0){8}}
\put(24,20){\framebox(4,4){$+$}}
 \put(28,20){\line(4,-3){4}}
 \put(32,14){\framebox(3,3){}}
\put(28,22){\line(1,0){8}}
\put(36,20){\framebox(4,4){$=$}}
\put(40,22){\line(1,0){8}}
\put(48,20){\framebox(4,4){$+$}}
 \put(52,20){\line(4,-3){4}}
 \put(56,14){\framebox(3,3){}}
\put(52,22){\line(1,0){8}}
\put(60,20){\framebox(4,4){$=$}}
\put(64,22){\line(1,0){8}}
\put(72,20){\framebox(4,4){$+$}}
 \put(76,20){\line(4,-3){4}}
 \put(80,14){\framebox(3,3){}}
\put(2,14){\line(0,1){6}}
\put(0,10){\framebox(4,4){$=$}}
\put(2,10){\line(0,-1){6}}
\put(26,14){\line(0,1){6}}
\put(24,10){\framebox(4,4){$=$}}
\put(26,10){\line(0,-1){6}}
\put(50,14){\line(0,1){6}}
\put(48,10){\framebox(4,4){$=$}}
\put(50,10){\line(0,-1){6}}
\put(74,14){\line(0,1){6}}
\put(72,10){\framebox(4,4){$=$}}
\put(74,10){\line(0,-1){6}}
\put(0,0){\framebox(4,4){$+$}}
 \put(4,0){\line(4,-3){4}}
 \put(8,-6){\framebox(3,3){}}
\put(4,2){\line(1,0){8}}
\put(12,0){\framebox(4,4){$=$}}
\put(16,2){\line(1,0){8}}
\put(24,0){\framebox(4,4){$+$}}
 \put(28,0){\line(4,-3){4}}
\put(32,-6){\framebox(3,3){}}
\put(28,2){\line(1,0){8}}
\put(36,0){\framebox(4,4){$=$}}
\put(40,2){\line(1,0){8}}
\put(48,0){\framebox(4,4){$+$}}
 \put(52,0){\line(4,-3){4}}
 \put(56,-6){\framebox(3,3){}}
\put(52,2){\line(1,0){8}}
\put(60,0){\framebox(4,4){$=$}}
\put(64,2){\line(1,0){8}}
\put(72,0){\framebox(4,4){$+$}}
 \put(76,0){\line(4,-3){4}}
 \put(80,-6){\framebox(3,3){}}
 
\put(8,63){\pos{bc}{$\tilde X_1$}}
\put(32,63){\pos{bc}{$\tilde X_2$}}

%%%
\put(14,64){\line(0,1){2}}
\put(38,64){\line(0,1){2}}
\put(62,64){\line(0,1){2}}
\put(12,66){\framebox(4,4){$$}}
\put(36,66){\framebox(4,4){$$}}
\put(60,66){\framebox(4,4){$$}}
\put(14,44){\line(0,1){2}}
\put(38,44){\line(0,1){2}}
\put(62,44){\line(0,1){2}}
\put(12,46){\framebox(4,4){$$}}
\put(36,46){\framebox(4,4){$$}}
\put(60,46){\framebox(4,4){$$}}
\put(14,24){\line(0,1){2}}
\put(38,24){\line(0,1){2}}
\put(62,24){\line(0,1){2}}
\put(12,26){\framebox(4,4){$$}}
\put(36,26){\framebox(4,4){$$}}
\put(60,26){\framebox(4,4){$$}}
\put(14,4){\line(0,1){2}}
\put(38,4){\line(0,1){2}}
\put(62,4){\line(0,1){2}}
\put(12,6){\framebox(4,4){$$}}
\put(36,6){\framebox(4,4){$$}}
\put(60,6){\framebox(4,4){$$}}
\put(0,52){\line(-1,0){2}}
\put(24,52){\line(-1,0){2}}
\put(48,52){\line(-1,0){2}}
\put(72,52){\line(-1,0){2}}
\put(-6,50){\framebox(4,4){$$}}
\put(18,50){\framebox(4,4){$$}}
\put(42,50){\framebox(4,4){$$}}
\put(66,50){\framebox(4,4){$$}}
\put(0,32){\line(-1,0){2}}
\put(24,32){\line(-1,0){2}}
\put(48,32){\line(-1,0){2}}
\put(72,32){\line(-1,0){2}}
\put(-6,30){\framebox(4,4){$$}}
\put(18,30){\framebox(4,4){$$}}
\put(42,30){\framebox(4,4){$$}}
\put(66,30){\framebox(4,4){$$}}
\put(0,12){\line(-1,0){2}}
\put(24,12){\line(-1,0){2}}
\put(48,12){\line(-1,0){2}}
\put(72,12){\line(-1,0){2}}
\put(-6,10){\framebox(4,4){$$}}
\put(18,10){\framebox(4,4){$$}}
\put(42,10){\framebox(4,4){$$}}
\put(66,10){\framebox(4,4){$$}}
\end{picture}
\vspace{2.5ex}
\caption{\label{fig:2DGridDM}%
Dual Forney factor graph of the 2D Ising model in an external 
field, where small 
boxes represent%factors as in
~(\ref{eqn:IsingKernelDual2}), unlabeled 
normal-size boxes 
represent%factors as in
~(\ref{eqn:IsingKernelDual}), 
and boxes containing 
$``+"$ symbols represent XOR factors, as in~(\ref{eqn:XOR}). 
%where $\oplus$ denotes addition in GF($2$).
\vspace{0.5mm}
}
\end{figure}
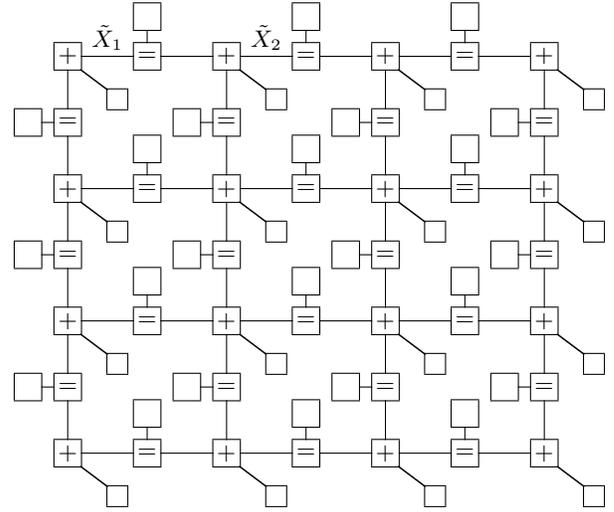

After suitable 
modifications, we can construct the dual Forney factor graph of the 2D Ising model, as shown in~\Fig{fig:2DGridDM}.
For binary variables $\tilde x_1, \tilde x_2, \ldots, \tilde x_k$ boxes containing 
$``+"$ symbols in~\Fig{fig:2DGridDM}, represent XOR factors as 
%where for binary variables $\tilde x_1, \tilde x_2, \ldots, \tilde x_k$, the XOR factors are defined as
\begin{equation} 
\label{eqn:XOR}
g(\tilde x_1, \tilde x_2, \ldots, \tilde x_k) =
[\tilde x_1 \oplus \tilde x_2 \oplus \ldots \oplus \tilde x_k=0]
\end{equation}
where $\oplus$ denotes the sum in GF($2$), the small boxes attached to each
XOR factor are as
\begin{equation} 
\label{eqn:IsingKernelDual2}
\lambda_{m}(\tilde x_m) = \left\{ \begin{array}{lr}
     \cosh H_{m}, & \text{if $\tilde x_m = 0$} \\
     -\sinh H_{m}, & \text{if $\tilde x_m = 1$} 
\end{array} \right.
\end{equation}
%and each factor~(\ref{eqn:IsingA}) is replaced by its 2D DFT,
%which has the following form 
%\begin{equation} 
%\label{eqn:IsingKernelD}
%\gamma_{k, \ell}(\tilde x_k, \tilde x_\ell) = \left\{ \begin{array}{ll}
%     4\cosh J_{k, \ell}, & \text{if $\tilde x_k = \tilde x_\ell = 0$} \\
%     4\sinh J_{k, \ell}, & \text{if $\tilde x_k = \tilde x_\ell = 1$} \\
%     0, & \text{otherwise.}
%  \end{array} \right.
%\end{equation}
%
%The corresponding dual Forney factor graph with factors as 
%in~(\ref{eqn:IsingKernelDual2}) and~(\ref{eqn:IsingKernelD})  
%is shown in~\Fig{fig:2DGridD}.%, see~\cite{MoLo:ISIT2013}.
and the unlabeled 
normal-size boxes attached to each equality constraint represent 
factors as 
\begin{equation} 
\label{eqn:IsingKernelDual}
\gamma_{k}(\tilde x_k) = \left\{ \begin{array}{ll}
     2\cosh J_{k}, & \text{if $\tilde x_k = 0$} \\
     2\sinh J_{k}, & \text{if $\tilde x_k = 1$}
  \end{array} \right.
\end{equation}

Here, $J_k$ is 
the coupling parameter associated with 
each bond. For more details, see~\cite{MoLo:ISIT2013, AY:2014}. % (the bond strength).

In this paper, we focus on ferromagnetic models, and as a result, all
the factors~(\ref{eqn:IsingKernelDual}) are positive.
In a 2D Ising model, the value of $Z$ is invariant under the
change of sign of the external field~\cite{Baxter07}. Therefore, without
loss of generality,  
we assume $H_m <  0$ for $1 \le m \le N$. With this assumption, all the factors~(\ref{eqn:IsingKernelDual2}) will also be positive. 
%\footnote{The factors in the dual domain can in general be negative or 
%complex-valued~\cite{MoLo:ITW2012}. Here, we require all factors to be positive
%because we need to define a probability 
%mass function in the dual factor graph, which is then used in the Monte Carlo methods of 
%Section~\ref{sec:IS}.
%Applying Monte Carlo methods to factor graphs with negative and complex 
%factors is discussed in~\cite{MoLo:ITW2012}.
% for Monte Carlo methods.
%Although ideas from~\cite{MoLo:ITW2012} can be applied to simulate models
%with negative or complex facotrs, in general, there might be issues with the
%sign problem~\cite{TW:05,LGS:90}.}.
%In Section~\ref{sec:IS}, we use the dual representation of the 2D Ising model 
%to give an alternative proof for the invariance 
%of $Z$ under the change of sign of the external field.

In the dual domain, 
we denote the partition function
by $Z_{\mathrm{d}}$. 
In the context of this paper, 
the normal factor graph duality 
theorem~\cite[Theorem 2]{AY:2011} states that
%\footnote{To be more precise,
%in our models there are no edges which are connected to only one node. 
%In Forney factor graphs such edges are represented by 
%a dongle symbol~\cite{Forney:01}. For
%the general form of the normal factor graph 
%duality theorem, see~\cite{Forney:11,AY:2011}.}
\begin{equation}
\label{eqn:NDual}
Z_{\mathrm{d}} = |\calX|^{N}Z
\end{equation}

%%%%%%%%%%%%%%%%%%%%%%%%%%%%%%%%%%%%%

Roughly speaking, the dual
representation transforms the low-temperature 
region (i.e., large J) to the high-temperature 
region (i.e., small J) and vice versa. Furthermore, due
to the presence of the XOR factors in~\Fig{fig:2DGridDM}, it is
possible to simulate a subset of the variables, followed by
doing computations on the remaining ones.
These properties can be employed to design efficient Monte Carlo methods in the dual domain to estimate $Z$ -- especially for  
cases that such an estimation might otherwise 
be difficult in the
original (primal) domain. 

In Section~\ref{sec:IS}, we design Monte Carlo methods in~\Fig{fig:2DGridDM} to estimate $Z_{\mathrm{d}}$, which can then be used to compute an estimate of
$Z$ via the normal factor 
graph duality theorem.
%\Fig{fig:2DGridDPotts} shows the corresponding dual Forney 
%factor graph with factors as 
%in~(\ref{eqn:PottsKernelDual}) and~(\ref{eqn:PottsKernelDual2}). 
%Note that compared 
%to~\Fig{fig:2DGridDM}, there are extra XOR factors next to 
%each equality constraint.

\section{Monte Carlo Methods}
\label{sec:IS}

We describe our Monte Carlo methods (importance sampling and uniform sampling) in the dual factor graph of the 2D Ising model
in an external %magnetic 
field. 

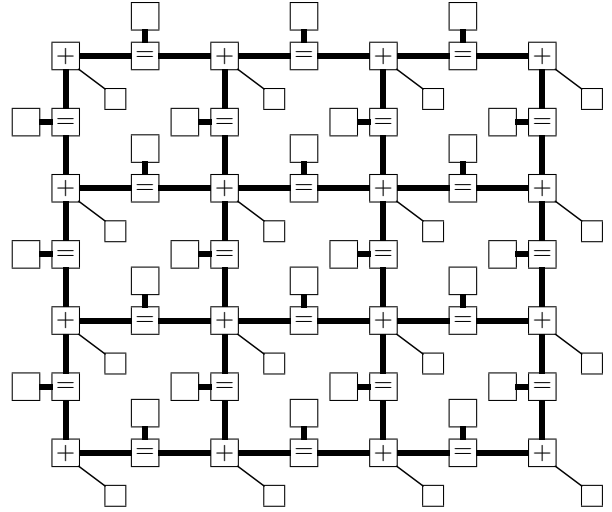
\begin{figure}[t]
\setlength{\unitlength}{0.88mm}
\centering
\begin{picture}(77,72.25)(0,0)
\small
\put(0,60){\framebox(4,4){$+$}}
 \put(8,54){\framebox(3,3){}}
\put(12,60){\framebox(4,4){$=$}}
\put(24,60){\framebox(4,4){$+$}}
 \put(32,54){\framebox(3,3){}}
\put(36,60){\framebox(4,4){$=$}}
\put(48,60){\framebox(4,4){$+$}}
 \put(56,54){\framebox(3,3){}}
\put(60,60){\framebox(4,4){$=$}}
\put(72,60){\framebox(4,4){$+$}}
 \put(80,54){\framebox(3,3){}}
 \put(0,50){\framebox(4,4){$=$}}
\put(24,50){\framebox(4,4){$=$}}
\put(48,50){\framebox(4,4){$=$}}
\put(72,50){\framebox(4,4){$=$}}
\put(0,40){\framebox(4,4){$+$}}
 \put(8,34){\framebox(3,3){}}
\put(12,40){\framebox(4,4){$=$}}
\put(24,40){\framebox(4,4){$+$}}
 \put(32,34){\framebox(3,3){}}
\put(36,40){\framebox(4,4){$=$}}
\put(48,40){\framebox(4,4){$+$}}
 \put(56,34){\framebox(3,3){}}
\put(60,40){\framebox(4,4){$=$}}
\put(72,40){\framebox(4,4){$+$}}
 \put(80,34){\framebox(3,3){}}
\put(0,30){\framebox(4,4){$=$}}
\put(24,30){\framebox(4,4){$=$}}
\put(48,30){\framebox(4,4){$=$}}
\put(72,30){\framebox(4,4){$=$}}
\put(0,20){\framebox(4,4){$+$}}
 \put(8,14){\framebox(3,3){}}
\put(12,20){\framebox(4,4){$=$}}
\put(24,20){\framebox(4,4){$+$}}
 \put(32,14){\framebox(3,3){}}
\put(36,20){\framebox(4,4){$=$}}
\put(48,20){\framebox(4,4){$+$}}
 \put(56,14){\framebox(3,3){}}
\put(60,20){\framebox(4,4){$=$}}
\put(72,20){\framebox(4,4){$+$}}
 \put(80,14){\framebox(3,3){}}
\put(0,10){\framebox(4,4){$=$}}
\put(24,10){\framebox(4,4){$=$}}
\put(48,10){\framebox(4,4){$=$}}
\put(72,10){\framebox(4,4){$=$}}
\put(0,0){\framebox(4,4){$+$}}
 \put(8,-6){\framebox(3,3){}}
\put(12,0){\framebox(4,4){$=$}}
\put(24,0){\framebox(4,4){$+$}}
\put(32,-6){\framebox(3,3){}}
\put(36,0){\framebox(4,4){$=$}}
\put(48,0){\framebox(4,4){$+$}}
 \put(56,-6){\framebox(3,3){}}
\put(60,0){\framebox(4,4){$=$}}
\put(72,0){\framebox(4,4){$+$}}
 \put(80,-6){\framebox(3,3){}}

%%%
\put(12,66){\framebox(4,4){$$}}
\put(36,66){\framebox(4,4){$$}}
\put(60,66){\framebox(4,4){$$}}
\put(12,46){\framebox(4,4){$$}}
\put(36,46){\framebox(4,4){$$}}
\put(60,46){\framebox(4,4){$$}}
\put(12,26){\framebox(4,4){$$}}
\put(36,26){\framebox(4,4){$$}}
\put(60,26){\framebox(4,4){$$}}
\put(12,6){\framebox(4,4){$$}}
\put(36,6){\framebox(4,4){$$}}
\put(60,6){\framebox(4,4){$$}}
\put(-6,50){\framebox(4,4){$$}}
\put(18,50){\framebox(4,4){$$}}
\put(42,50){\framebox(4,4){$$}}
\put(66,50){\framebox(4,4){$$}}
\put(-6,30){\framebox(4,4){$$}}
\put(18,30){\framebox(4,4){$$}}
\put(42,30){\framebox(4,4){$$}}
\put(66,30){\framebox(4,4){$$}}
\put(-6,10){\framebox(4,4){$$}}
\put(18,10){\framebox(4,4){$$}}
\put(42,10){\framebox(4,4){$$}}
\put(66,10){\framebox(4,4){$$}}

 \put(4,60){\line(4,-3){4}}
 \put(28,60){\line(4,-3){4}}
 \put(52,60){\line(4,-3){4}}
 \put(76,60){\line(4,-3){4}}

 \put(4,40){\line(4,-3){4}}
 \put(28,40){\line(4,-3){4}}
 \put(52,40){\line(4,-3){4}}
 \put(76,40){\line(4,-3){4}}
 \put(4,20){\line(4,-3){4}}
 \put(28,20){\line(4,-3){4}}
 \put(52,20){\line(4,-3){4}}
 \put(76,20){\line(4,-3){4}}
 \put(4,0){\line(4,-3){4}}
 \put(28,0){\line(4,-3){4}}
 \put(52,0){\line(4,-3){4}}
 \put(76,0){\line(4,-3){4}}

%%%%%%%%%%%%%%%%%
 \linethickness{0.67mm}
 %%%%%%%%%%%%%%%%%
\put(14,4){\line(0,1){2}}
\put(38,4){\line(0,1){2}}
\put(62,4){\line(0,1){2}} 

 \put(14,64){\line(0,1){2}}
\put(38,64){\line(0,1){2}}
\put(62,64){\line(0,1){2}}

\put(14,44){\line(0,1){2}}
\put(38,44){\line(0,1){2}}
\put(62,44){\line(0,1){2}}

 \put(14,24){\line(0,1){2}}
\put(38,24){\line(0,1){2}}
\put(62,24){\line(0,1){2}}

\put(0,52){\line(-1,0){2}}
\put(24,52){\line(-1,0){2}}
\put(48,52){\line(-1,0){2}}
\put(72,52){\line(-1,0){2}}

\put(0,32){\line(-1,0){2}}
\put(24,32){\line(-1,0){2}}
\put(48,32){\line(-1,0){2}}
\put(72,32){\line(-1,0){2}}

\put(0,12){\line(-1,0){2}}
\put(24,12){\line(-1,0){2}}
\put(48,12){\line(-1,0){2}}
\put(72,12){\line(-1,0){2}}

 %%%%%%%%%%%%%%%%%%
 
\put(4,62){\line(1,0){8}}        
\put(16,62){\line(1,0){8}}
\put(28,62){\line(1,0){8}}       
\put(40,62){\line(1,0){8}}
\put(52,62){\line(1,0){8}}       
\put(64,62){\line(1,0){8}}
\put(2,54){\line(0,1){6}}
\put(2,50){\line(0,-1){6}}
\put(26,54){\line(0,1){6}}
\put(26,50){\line(0,-1){6}}
\put(50,54){\line(0,1){6}}
\put(50,50){\line(0,-1){6}}
\put(74,54){\line(0,1){6}}
\put(74,50){\line(0,-1){6}}
\put(4,42){\line(1,0){8}}
\put(16,42){\line(1,0){8}}
\put(28,42){\line(1,0){8}}
\put(40,42){\line(1,0){8}}
\put(52,42){\line(1,0){8}}
\put(64,42){\line(1,0){8}}
\put(2,34){\line(0,1){6}}
\put(2,30){\line(0,-1){6}}
\put(26,34){\line(0,1){6}}
\put(26,30){\line(0,-1){6}}
\put(50,34){\line(0,1){6}}
\put(50,30){\line(0,-1){6}}
\put(74,34){\line(0,1){6}}
\put(74,30){\line(0,-1){6}}
\put(4,22){\line(1,0){8}}
\put(16,22){\line(1,0){8}}
\put(28,22){\line(1,0){8}}
\put(40,22){\line(1,0){8}}
\put(52,22){\line(1,0){8}}
\put(64,22){\line(1,0){8}}
\put(2,14){\line(0,1){6}}
\put(2,10){\line(0,-1){6}}
\put(26,14){\line(0,1){6}}
\put(26,10){\line(0,-1){6}}
\put(50,14){\line(0,1){6}}
\put(50,10){\line(0,-1){6}}
\put(74,14){\line(0,1){6}}
\put(74,10){\line(0,-1){6}}
\put(4,2){\line(1,0){8}}
\put(16,2){\line(1,0){8}}
\put(28,2){\line(1,0){8}}
\put(40,2){\line(1,0){8}}
\put(52,2){\line(1,0){8}}
\put(64,2){\line(1,0){8}}
 
%\put(8,63){\pos{bc}{$\tilde X_1$}}
%\put(32,63){\pos{bc}{$\tilde X_2$}}
%
\end{picture}
\vspace{2.5ex}
\caption{\label{fig:2DGridDPart}%
A partitioning of the variables in Fig.~\ref{fig:2DGridDM}. The thick edges (bonds) represent $\tilde \X_A$ and
edges connected to the unlabeled 
small boxes represent $\tilde \X_B$. Here, $\tilde \X_B$ is
a linear combination (involving XOR factors) of $\tilde \X_A$.
%\vspace{-1.85mm}
}
\end{figure}

In~\Fig{fig:2DGridDM}, let us partition %the set of random variables 
$\tilde \X$, into $\tilde \X_A$ and $\tilde \X_B$, with the restriction that $\tilde \X_B$ is a
linear combination (involving the XOR factors) of
$\tilde \X_A$. % see~\cite{MoLo:ISIT2013}. 
An example of such a partitioning is illustrated 
in Fig.~\ref{fig:2DGridDPart}, where $\tilde \X_B$
is the
set of all the edges connected to 
the small unlabeled boxes (which are involved in 
factors~(\ref{eqn:IsingKernelDual2})), and $\tilde \X_A$
is the set of all the bonds (which are involved in 
factors~(\ref{eqn:IsingKernelDual}) and are marked by thick edges). %in~\Fig{fig:2DGridDPart}.
As will be discussed, this choice of partitioning is appropriate for models in a strong external field. 

In this set-up, a valid configuration $\tilde \x = (\tilde \x_A, \tilde \x_B)$ in the dual factor graph can be created by assigning values to $\tilde \X_A$, followed by updating
$\tilde \X_B$ as a linear combination of $\tilde \X_A$.

%It also simplifies the notations in the sequel.

Accordingly, let us define
\begin{IEEEeqnarray}{r,C,l}
\Gamma(\tilde \x_A) & \eqdef & 
\prod_{\tilde x_k \in \tilde \x_A} \gamma _{k}(\tilde x_k) \label{eqn:PartG}\\
\Lambda(\tilde \x_B) & \eqdef & 
\prod_{\tilde x_m \in \tilde \x_B} \lambda_{m}(\tilde x_m)\label{eqn:PartL}
\end{IEEEeqnarray}

From~(\ref{eqn:PartG}), we define the following probability 
mass function in $\calX^{|\B|}$
\begin{IEEEeqnarray}{r,C,l}
\label{eqn:AuxDist}
q(\tilde \x_A) & \eqdef & \frac{\Gamma(\tilde \x_A)}{Z_q}, \; \qquad \forall \tilde \x_A \in \calX^{|\B|}
\end{IEEEeqnarray}

The probability 
mass function~(\ref{eqn:AuxDist})
%$q(\tilde \x_A)$ in
has two key properties.
First, its partition function $Z_q$ is analytically available as
\begin{IEEEeqnarray}{r,C,l}
Z_q  & = & \sum_{\tilde \x_A} \Gamma(\tilde \x_A) \\
        & = & \prod_{k \in \B} 2(\cosh J_k + \sinh J_k) \\
        & = & 2^{|\B|} \text{exp}\big(\sum_{k \in \B} J_k\big) \label{eqn:Zq} 
        %4^{|\B|} e^{\sum_{k = 1}^{|\B|} J_k}
\end{IEEEeqnarray}
where $|\B|$ denotes the cardinality
of $\B$,
%where $|\B|$ is the cardinality
%of $\B$, 
which is equal to the number of bonds in the lattice (cf.\ Section \ref{sec:Ising}). 

%In our numerical experiments in
%Section~\ref{sec:Num}, we consider 2D models with periodic 
%boundary conditions, in this case $|\B| = 2N$.
%The value
%of $Z_q$ is thus a function of the sum of all the
%coupling parameters.
%set of all the unordered 
%pairs with non-zero interactions (cf.\ Section \ref{sec:Ising}).

%If the model is at very low temperature, 
%i.e., $J_k$ is very large, the above algorithm will be 
%similar to
%uniform sampling, where each 
%sample $x_k^{(\ell)}$ is drawn uniformly and independently 
%from $\calX$. % see~(\ref{eqn:ISL}).
%%$\frac{\gamma_k(0)}{\gamma_k(0) + \gamma_k(1)}$.

Second, it is straightforward to draw 
\emph{independent} samples %``independent" samples 
$\tilde \x_A^{(1)}, \tilde \x_A^{(2)}, \ldots, \tilde \x_A^{(\ell)}, \ldots$,
according to $q(\tilde \x_A)$. The product form of~(\ref{eqn:PartG}) 
indicates that to draw $\tilde \x_A^{(\ell)}$ we can do the following.

\vspace{0.3ex}

\begin{itemize}
\itemsep1.7pt
%\item[] \rule{81.4mm}{1.6pt}
\item[] \emph{draw} $u_1^{(\ell)}, u_2^{(\ell)}, \ldots, u_{|\B|}^{(\ell)}\overset{\text{i.i.d.}}{\sim} \, \mathcal{U}[0,1]$
\item[] {\bf for} $k = 1$ {\bf to} $|\B|$
%\item[] \hspace{4.5mm} draw $u_m^{(\ell)} {\sim} \,\mathcal{U}[0,1]$ 
\item [] \hspace{4.5mm} {\bf if} $u_k^{(\ell)} < \frac{1}{2}(1+e^{-2J_k})$
\item [] \hspace{10mm}  $\tilde x_{A,k}^{(\ell)} = 0$
\item [] \hspace{4.5mm} {\bf else}
\item [] \hspace{10mm}  $\tilde x_{A,k}^{(\ell)} = 1$
\item [] \hspace{4.5mm} {\bf end if}
\item[] {\bf end for}
%\hfill$\blacksquare$
%\item[] \rule{81.4mm}{1.3pt}
\end{itemize}
\vspace{0.3ex}
The quantity
$\frac{1}{2}(1+e^{-2J_k})$ is equal to 
$\gamma_k(0)/\big(\gamma_k(0) + \gamma_k(1)\big)$.

\vspace{0.2ex}

As $\tilde \x_B$ is a 
linear combination 
%(involving the XOR factors) 
of $\tilde \x_A$, updating $\tilde \x_B^{(\ell)}$
is easy after generating $\tilde \x_A^{(\ell)}$. These samples are then used in the following 
importance sampling algorithm to estimate $Z_\text{d}$.
\vspace{0.3ex}
\begin{itemize}
\itemsep1.7pt
%\item[] \rule{81.4mm}{1.6pt}
\item[] {\bf for} $\ell = 1$ {\bf to} $L$
\item[] \hspace{4.5mm} \emph{draw} $\x_A^{(\ell)}$ \emph{according to} $q(\tilde \x_A)$
\item[] \hspace{4.5mm} \emph{update} $\tilde \x_B^{(\ell)}$
\item[] {\bf end for}
\item[] \emph{compute}
\begin{IEEEeqnarray}{r,C,l}
\label{eqn:EstR}
\hat Z_{\text{IS}} & = & \frac{Z_q}{L} \sum_{\ell = 1}^L \Lambda(\tilde \x_B^{(\ell)}) 
\end{IEEEeqnarray}
\end{itemize}
%\vspace{0.3ex}
%(Note that $Z_q$ is analytically available in~(\ref{eqn:Zq})).

It follows that, $\hat Z_{\text{IS}}$ is an unbiased estimator of $Z_\text{d}$. 
%Indeed, $\E[\, \hat Z_{\text{IS}}\,] = Z_\text{d}$.
Indeed
\begin{equation}
\E[\, \hat Z_{\text{IS}}\,] = Z_\text{d}
\end{equation}

%%%%%%%%%%%%%%%%%%%%%%%%%%%%%%%%%%%%%

The proposed
importance sampling scheme 
can yield an estimate of $Z_\text{d}$, which can then be used to estimate
$Z$ in~(\ref{eqn:PartFunction}), using the normal factor 
graph duality theorem (cf.\ Section~\ref{sec:NFGD}).

The accuracy of~(\ref{eqn:EstR}) depends on the
fluctuations of $\Lambda(\tilde \x_B)$. If $\Lambda(\tilde \x_B)$ varies 
smoothly, $\hat Z_{\text{IS}}$ will
have a small variance. With our choice of 
partitioning in~(\ref{eqn:PartG}) and~(\ref{eqn:PartL}),
we expect to observe a small variance 
if the model is in a strong (negative) external 
field. See Appendix~I for a discussion.

%E.g., for an Ising
%model in a weak external field but with a mixture of weak and strong coupling
%parameters, it seems plausible that choosing $\tilde \X_B$ to be the set of
%variables involved in factors with strong couplings, yields an estimator
%with a smaller variance.

%For models in a weak external field, the efficiency of the importance sampling algorithm 
%on the dual factor graph should be 
%compared to the efficiency of Monte Carlo methods applied directly to the original factor 
%graph, as in Figs.~\ref{fig:2DGrid} and~\ref{fig:2DGridOrig}.
%For model at hight temperature and in a weak external field, it is more
%efficient to
%perform 
%Monte Carlo methods directly on the original factor 
%graph as in \Fig{fig:2DGrid}.

%Note that, the Hamming weight~\cite{RJM:77} of $\tilde \x_B^{(\ell)}$ 
%is even, as
%\begin{equation}
%\bigoplus_{k=1}^N \tilde x_{B,k}^{(\ell)} = 
%\bigoplus_{k=1}^{|\B|}\tilde x_{A, k}^{(\ell)} \oplus \tilde x_{A,k}^{(\ell)} = 0
%\end{equation} 
%Therefore, $\Lambda(\tilde \x_B^{(\ell)})$
%will always take positive values even if $H_m$ in (\ref{eqn:IsingKernelDual2}) is
%positive.
%However, setting $H_m < 0$ makes the formulation of the problem more clear.
%This can also be considered as a simple proof that $Z$ is invariant under the
%change of sign of the external field (cf.\ Section~\ref{sec:NFGD}).
%However, setting $H_m < 0$ makes the formulation more clear.

We can design a uniform sampling algorithm
by drawing $\x_{A}^{(\ell)}$ uniformly and independently 
from $\calX^{|\B|}$, and by applying
\begin{IEEEeqnarray}{r,C,l}
\label{eqn:EstUn}
\hat Z_{\text{Unif}} & = & \frac{|\calX|^{|\B|}}{L} \sum_{\ell = 1}^L \Gamma(\tilde \x_A^{(\ell)})\Lambda(\tilde \x_B^{(\ell)}) 
\end{IEEEeqnarray}

It is easy to verify that, $\E[\, \hat Z_{\text{Unif}}\,] = Z_\text{d}$.
%\begin{equation} 
%\label{eqn:EstU}
%\hat Z_{\text{Unif}} = 
%\frac{|\calX|^{|\B|}}{L} \sum_{\ell = 1}^L \Gamma(\tilde \x_A^{(\ell)})\Lambda(\tilde \x_B^{(\ell)})
%\end{equation}
%
%It is easy to verify that, $\E[\, \hat Z_{\text{Unif}}\,] = Z_\text{d}$.
%\vspace{0.3mm}
%It must be emphasized that, the cost of generating $\tilde \x_{A}^{(\ell)}$ using the importance sampling 
%scheme (i.e., drawing an independent sample according to~(\ref{eqn:AuxDist})) is virtually the same as the cost of
%generating a sample with uniform sampling (i.e., drawing a sample uniformly and independently in $\calX^N$).

The efficiency of the uniform sampling 
and the importance sampling algorithms will be close if $J_k$ is very 
large (i.e., when the model is 
at very low temperature). However,
for a wider range of parameters, importance 
sampling outperforms uniform sampling -- as will be illustrated in our numerical 
experiments in Section~\ref{sec:Num}.

%; see Appendix I.
%Applying uniform sampling and Gibbs sampling in the dual domain
%to 2D Ising models in the absence of an external field are discussed in~\cite{MoLo:ISIT2013}.

%Finally, note that in~(\ref{eqn:ISL}) by drawing each 
%sample $x_k^{(\ell)}$ uniformly and independently from $\calX$, we can
%introduce a uniform sampling scheme.
%The efficiency of such a uniform sampling scheme is similar to the 
%efficiency of the importance sampling scheme for models at very low temperature
%(very large $|J_k|$). 
%However, for a wider range of model parameters, uniform sampling
%does not perform well, as will be illustrated in our numerical 
%experiments in Section~\ref{sec:Num}.

If the model is in a relatively strong external field, 
we can consider
applying annealed importance sampling~\cite{NealIS:2001}; see Appendix II. 
%see~(\ref{eqn:IsingKernelDual2}).
%see numerical experiments in Section~\ref{sec:Num}.
The choice of partitioning in the dual graph is 
%completely 
arbitrary, as long as $\tilde \x_B$ can be computed as linear combinations
of $\tilde \x_A$. The partitioning
in~\Fig{fig:2DGridDPart} is suitable for models in a strong 
external field. An example of a partitioning suitable for models with
strong couplings is described in~\cite{MeMo:2014b}.

Finally, note that a good general strategy to reduce the variance of Monte Carlo methods in~\Fig{fig:2DGridDM}, is to include factors with larger
model parameters (coupling parameters $J$ and the external magnetic field $H$) in $\Lambda(\tilde \x_B)$.

\section{Numerical Experiments}
\label{sec:Num}

We apply the proposed Monte Carlo methods of Section~\ref{sec:IS}
to estimate the log partition function per 
site, i.e., $\frac{1}{N}\ln Z$, of the 
2D ferromagnetic Ising model in an external field with spatially varying model parameters.
%We recall that $Z$ is invariant under the
%change of sign of the field;
%we set $H_m <  0$ to make all the factors 
%as in~(\ref{eqn:IsingKernelDual2}) positive.

%%%%%%%%%%%%%%%%%%%%%%%
%These extra edges are not
%plotted in Figs.~\ref{fig:2DGrid} to~\ref{fig:2DGridPotts} to avoid clutter.

%In this case, the size of the state space for Monte Carlo
%methods only depends on the size of $\tilde X_A$.

All simulation results show $\frac{1}{N}\ln Z$ vs.\ the number
of samples for one instance of the Ising model of 
size $N = 30\times 30$ and with periodic
boundary conditions.
%where to create periodic boundary conditions
%we need to add extra edges %(with appropriate factors) 
%to connect the 
%sites on opposite sides of the boundary. 
In this case $|\B| = 2N$.

\begin{figure}[t!]
\centering
\includegraphics[width=0.96\linewidth]{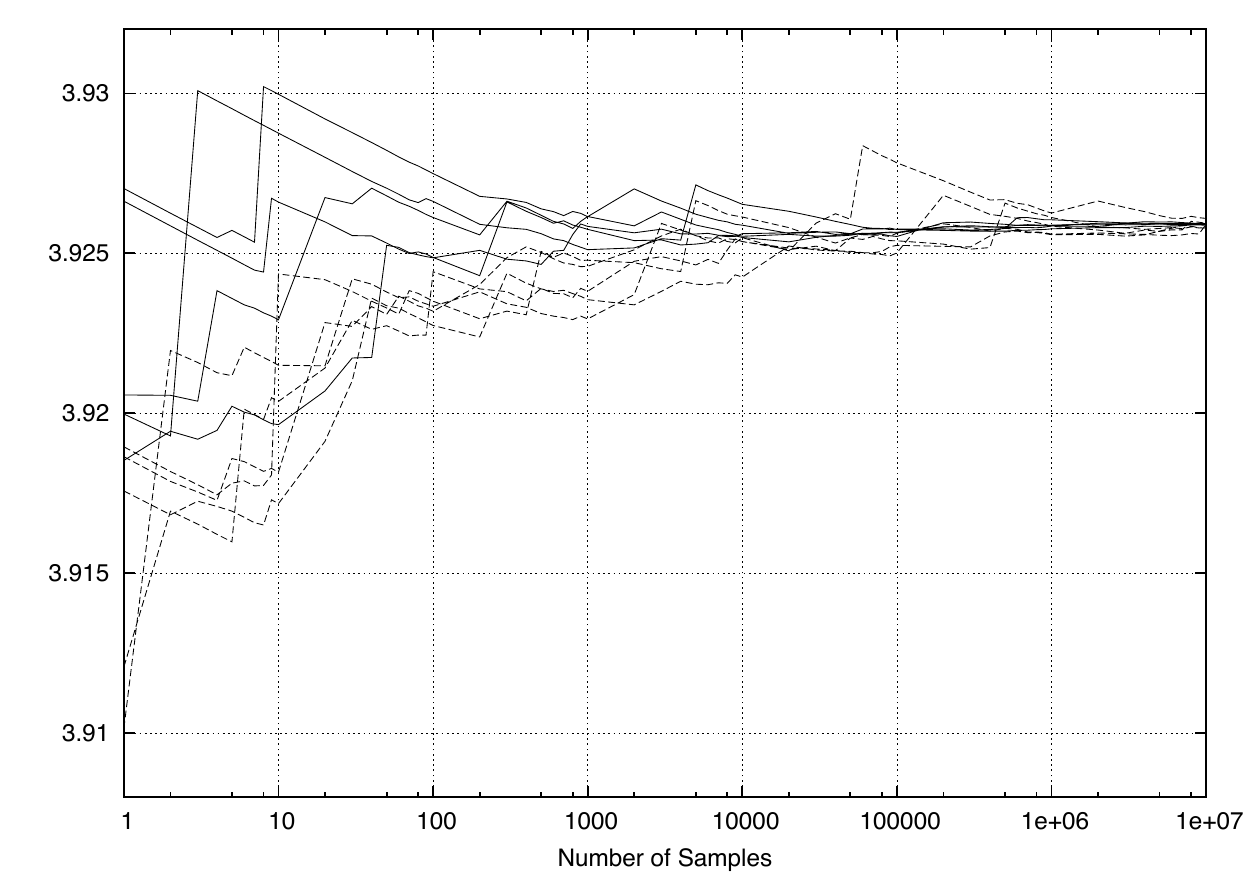}
\caption{\label{fig:FerIsing1}%
Estimated $\ln(Z)$ per site vs.\ the number of samples
for a $30\times 30$ Ising model, with
$J\sim\calU[1.3, 1.5]$ and $H\sim\calU[-1.25, -1.0]$. 
The plot shows five different sample paths obtained from importance sampling (solid lines) and five different sample paths obtained from uniform sampling (dashed lines)
in the dual factor graph.}
\vspace{1.5ex}
%\end{figure}
%\begin{figure}[t!]
\includegraphics[width=0.96\linewidth]{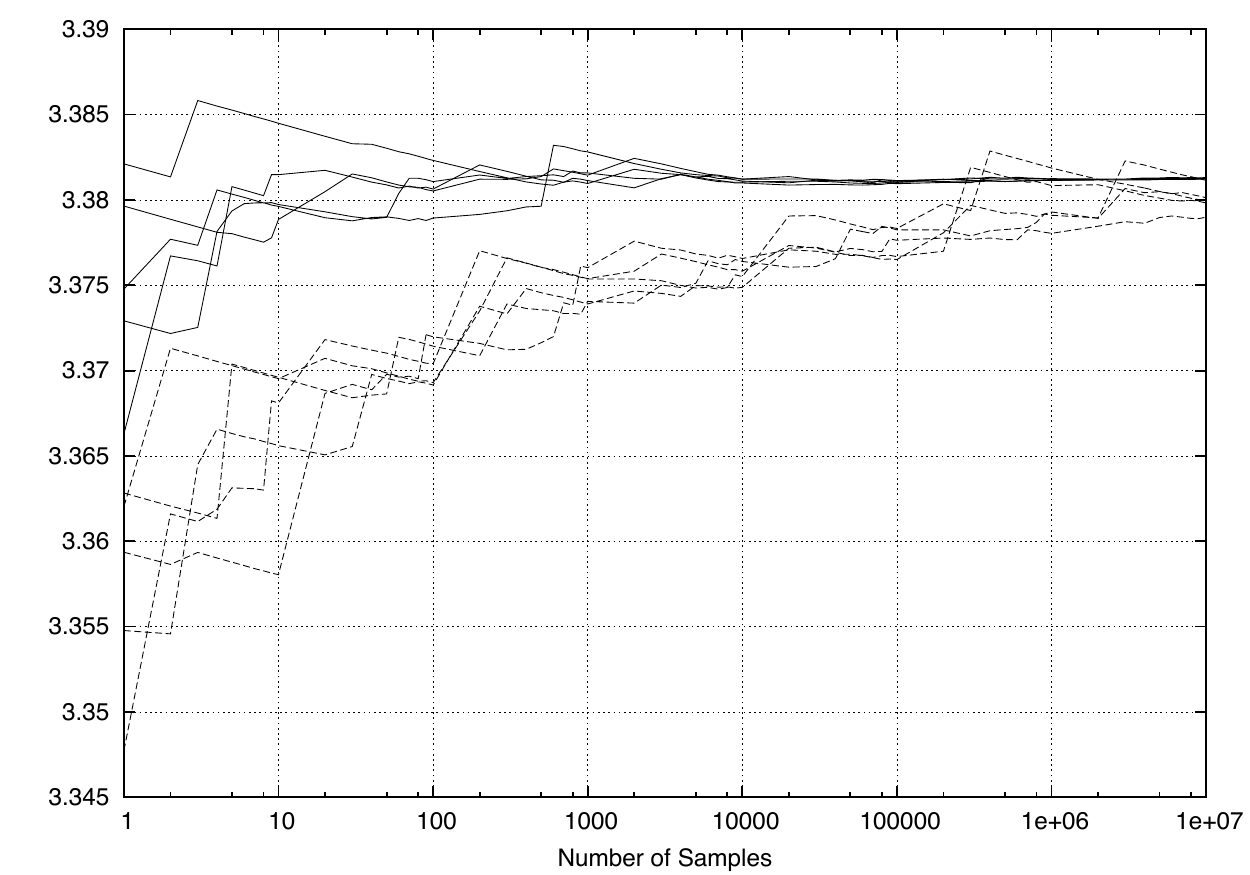}
\caption{\label{fig:FerIsing2}%
Everything as in Fig.~\ref{fig:FerIsing1}; but with $J\sim\calU[0.75, 1.5]$.}
\end{figure}

In our first two experiments we
set $H_{m} \overset{\text{i.i.d.}}{\sim} \calU[-1.25, -1.0]$.
The coupling parameters 
are set to $J_{k} \overset{\text{i.i.d.}}{\sim} \calU[1.3, 1.5]$ 
in the first experiment and to $J_{k} \overset{\text{i.i.d.}}{\sim} \calU[0.75, 1.5]$ 
in the second experiment. 
Simulation results obtained from importance 
sampling (solid lines) and uniform sampling (dashed lines) %in the dual factor graph 
are shown in Figs.~\ref{fig:FerIsing1} and \ref{fig:FerIsing2}.
%From Figs.~\ref{fig:FerIsing1} and \ref{fig:FerIsing2},
The estimated log partition functions per site are about $3.926$ and $3.381$, respectively.

%%%%%%%%%%%%%%%%%%%%%%%%%%%%%%%%%%%%%%%
%%%%%%%%%%%%%%%%%%%%%%%%%%%%%%%%%%%%%%%%%%%%%

For very large coupling parameters, 
(corresponding to models at very low temperature), 
convergence of uniform sampling is comparable to the 
convergence of the importance
sampling algorithm (see~\Fig{fig:FerIsing1}). 
However, as we observe in Fig~\ref{fig:FerIsing2}, uniform sampling has issues 
with slow convergence 
for a wider range of coupling parameters,
while the importance sampling algorithm performs
well in all the ranges.

\begin{figure}[t!]
\includegraphics[width=0.96\linewidth]{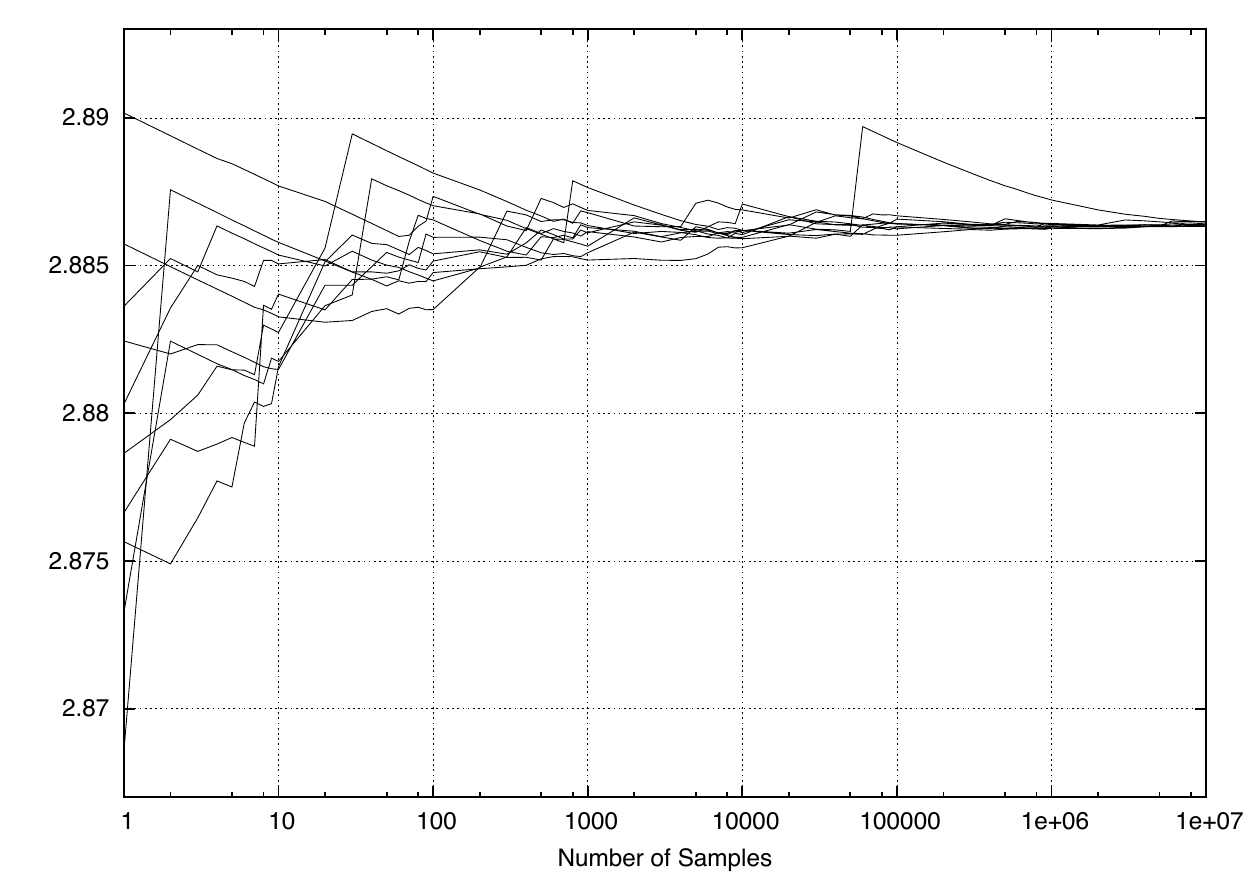}
\caption{\label{fig:FerIsing3}%
Estimated $\ln(Z)$ per site vs.\ the number of samples
for a $30\times 30$ Ising model, with $J\sim\calU[0.25, 1.5]$ and $H\sim\calU[-1.25, -1.0]$. 
The plot shows ten different sample paths obtained from importance sampling
in the dual factor graph.}
\vspace{1.5ex}
\includegraphics[width=0.96\linewidth]{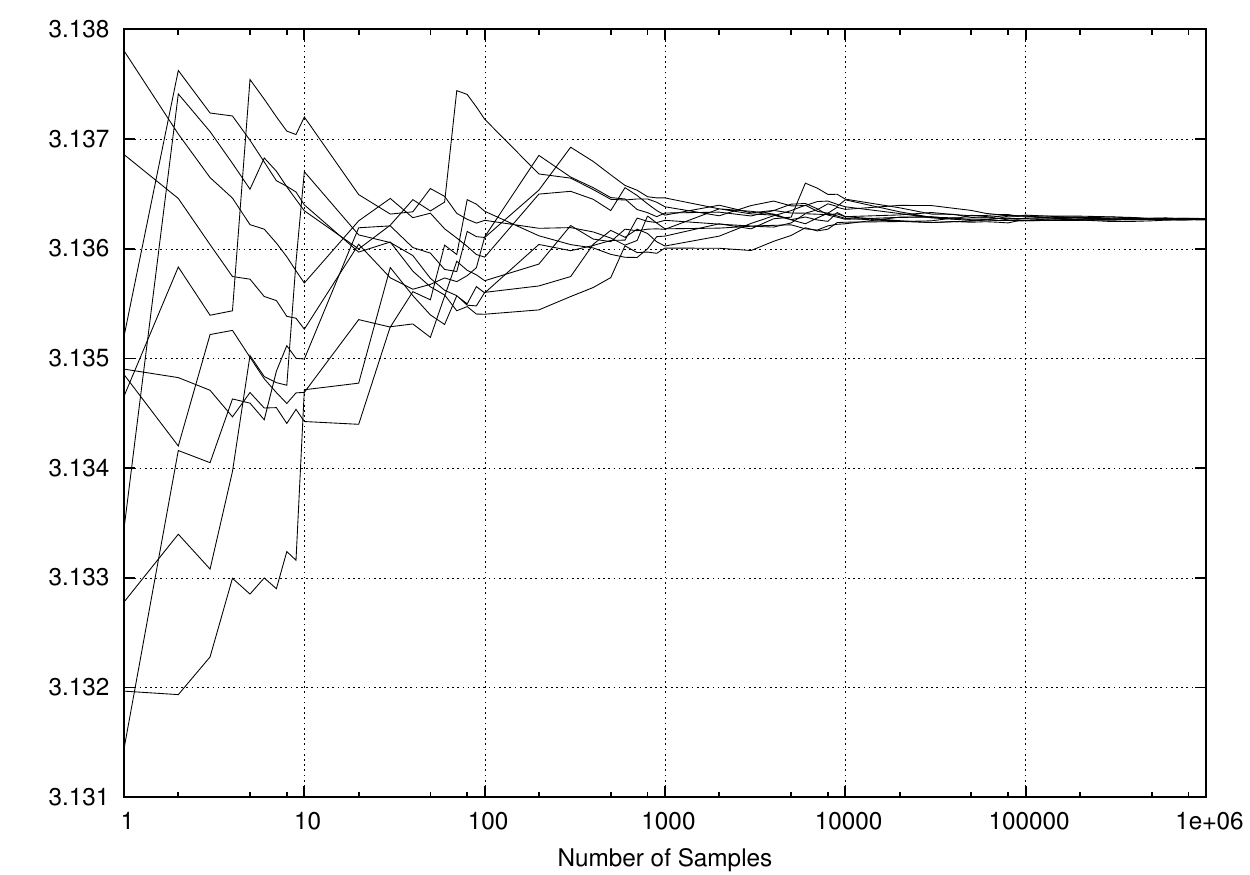}
\caption{\label{fig:FerIsing44}%
Everything as in Fig.~\ref{fig:FerIsing3}; but with $H \sim\calU[-1.5, -1.25]$.}
\end{figure}

In our last two experiments we
set $J_{k} \overset{\text{i.i.d.}}{\sim} \calU[0.25, 1.5]$.
In the third experiment, we set $H_{m} \overset{\text{i.i.d.}}{\sim} \calU[-1.25, -1.0]$.
Fig.~\ref{fig:FerIsing3} shows simulation
results obtained from importance sampling, where the estimated log partition function per site 
is about $2.886$. We set $H_{m} \overset{\text{i.i.d.}}{\sim} \calU[-1.5, -1.25]$ in the last
experiment. The estimated $\frac{1}{N}\ln Z$ from~Fig.~\ref{fig:FerIsing44} is 
about $3.1362$. We observe that convergence of the importance sampling
algorithm improves as $|H|$ becomes 
larger (see Appendix~I).

%%%%%%%%%%%%%%%%%%%%%%%%%%%%%%%%%%%%%%%%

%%%%%%%%%%%%%%%%%%%%%%%%%%%%%%%%%%%%%%%%%%%%

\section{Conclusion}

Monte Carlo methods were proposed in the dual Forney factor graph 
to estimate the 
partition function of the 2D ferromagnetic Ising model in an 
external magnetic field.
We described a method to partition the variables in the dual factor graph and
introduced an auxiliary probability mass function accordingly. 

The
methods can efficiently estimate the partition function in a wide range 
of model parameters; in particular (with our choice of partitioning), when 
the Ising model is in a strong external magnetic field. 
Indeed, convergence of the methods
improve as the external field becomes stronger.
Depending on the values and the spatial distribution of 
the model parameters, 
%(coupling parameters and the external magnetic field) 
different partitionings yield schemes with different %dynamics.
convergence properties. Generalizations of the proposed methods to the 
$q$-state Potts model are discussed 
in~\cite[Section V]{MeMo:2014a}. Comparisons 
with deterministic algorithms in the primal domain (e.g., the generalized belief propagation and the tree expectation propagation algorithms, as 
done in~\cite{Gomez:2010}) are left for future work.
%which is an interesting topic to explore. 
%Depending on the values of the model parameters 
%%(coupling parameters and the external magnetic field) 
%and their spatial distributions,
%different choices of partitioning will yield schemes with different dynamics. 
\section*{Acknowledgements}

%\vspace{0.5mm}
 
The author would like to thank Hans-Andrea Loeliger, Pascal Vontobel, David Forney, Justin Dauwels, and 
Ali Al-Bashabsheh for their %many helpful %many helpful
comments that greatly improved the presentation of this paper. The author gratefully acknowledges 
the support of Albert Guill\'{e}n i F\`{a}bregas at UPF.

%The author would like to thank H.-A. Loeliger, P.~Vontobel, D.~Forney, J.~Dauwels, and 
%A.~Al-Bashabsheh for their %many helpful %many helpful
%comments that greatly improved the presentation of this paper. 
%The author gratefully acknowledges 
%the support of Albert Guill\'{e}n i F\`{a}bregas at UPF.
%%The author gratefully acknowledges 
%%the support of Albert Guill\'{e}n i F\`{a}bregas at UPF.
%%The author also wishes to thank Justin Dauwels for his comments
%%on an earlier draft of this paper. 
%%

%\vspace{1.5mm}
\section*{Appendix~I\\ Convergence of Monte Carlo methods 
in the Dual Forney Factor graph}

For simplicity, we assume that the coupling parameter and the external field are both constant, 
denoted by $J$ and $|H|$, respectively.
In the dual factor graph, let us replace each factor~(10) by
\begin{equation} 
\label{eqn:IsingKerneltanh1}
\lambda(\tilde x_m) = (\tanh |H|)^{\tilde x_m} 
\end{equation}
and each factor~(12) by 
\begin{equation} 
\label{eqn:IsingKerneltanh2}
\gamma(\tilde x_k) = (\tanh J)^{\tilde x_k}
\end{equation}
%and
%\begin{equation} 
%\label{eqn:Zdd}
%\frac{Z_\text{d}}{Z_{d^{'}}} = 2^{N+2|\B|}\prod_{m=1}^N \cosh H_{m} \prod_{k \in \B} \cosh J_{k}
%\end{equation}

The required scale factor $S$ to recover $Z_\text{d}$ can be easily computed by multiplying all 
the local scale factors as 
\begin{equation} 
\label{eqn:DDRatio}
S = (2\cosh J)^{|\B|}(\cosh |H|)^N
\end{equation}

Note that, $\lim_{t \to \infty} \tanh t = 1$, therefore
in a strong external field (i.e., large $|H|$) and at low temperature 
(i.e., large $J$), $\tanh |H|$
and $\tanh J$ both tend to constant, which gives reasons 
for the fast convergence of uniform sampling in this case. 
%In the importance sampling algorithm, independent samples 
%are drawn according to $q(\tilde \x_A)$ in~(\ref{eqn:AuxDist}), thus the only 
%requirement to achieve fast convergence is to have a strong
%external field. 
Indeed, convergence of the uniform sampling 
algorithm in the dual domain improves as $J$ and $|H|$ both become larger. And for a fixed $J$, convergence 
of the importance sampling algorithm improves as $|H|$ 
becomes larger. For more details, 
see~\cite[Appendix~I]{MeMo:2014a}.

\section*{Appendix II\\ Annealed Importance Sampling in the Dual Forney Factor Graph}

We briefly explain how to employ annealed importance sampling~\cite{NealIS:2001} in the dual factor 
graph to estimate the partition function of the 2D Ising model, when the model is in a relatively strong consistent 
external field.

Again, for simplicity, we assume that the coupling parameter and the external field are both constant. 
The partition function is thus denoted by $Z_{\text{d}}(J, |H|)$. 
We express $Z_{\text{d}}(J, |H|)$ using a sequence of intermediate partition functions %, which is needed for annealed importance sampling, 
by varying $|H|$ in $V$ levels as
\begin{equation}
\label{eqn:AIS}
Z_{\text{d}}(J, |H|) = Z_{\text{d}}(J, |H|^{\alpha_V})\prod_{v = 0}^{V-1} 
\frac{Z_{\text{d}}(J, |H|^{\alpha_v})}{Z_{\text{d}}(J, |H|^{\alpha_{v+1}})}
%\frac{Z(J, H^{\alpha_0})}{Z(J, H^{\alpha_1})}\frac{Z(J, H^{\alpha_1})}{Z(J, H^{\alpha_2})}\cdots\frac{Z(J, H^{\alpha_{M-1}})}{Z(J, H^{\alpha_M})}
\end{equation}

Here, unlike typical annealing strategies applied in 
the original domain, 
$(\alpha_0, \alpha_1, \ldots, \alpha_V)$ is an 
increasing 
sequence, with $1 = \alpha_0 < \alpha_1 < \cdots < \alpha_V$.

If $\alpha_V$ is large enough, $Z_{\text{d}}(J, |H|^{\alpha_V})$ can be estimated efficiently via our proposed 
Monte Carlo methods. As for the intermediate steps, a sampling 
technique that leaves the target distribution invariant (e.g., Metropolis-Hastings algorithms or Gibbs sampling~\cite{Neal:proinf1993r}) is 
required at each level. 
%These intermediate target probability distributions correspond to
%the intermediate partition functions.
%\vspace{3mm}
%
%\begin{itemize}
%\itemsep4.5pt
%%\item[] \rule{81.4mm}{1.6pt}
%\item[] \hspace{-4.2mm}\emph{set} $s = \tilde x_{A,k}^{(\ell-1)} \oplus 1$
%\item[] \hspace{-4.2mm}\emph{draw} $u_k^{(\ell)} \sim \, \mathcal{U}[0,1]$
%%\item[] \hspace{4.5mm} draw $u_m^{(\ell)} {\sim} \,\mathcal{U}[0,1]$ 
%%\item []  {\bf if} $u_k^{(\ell)}  < \frac{1}{1 \, + \, (\tanh J)^{1-2s} (\tanh H^{\alpha_v})^{2(1- \tilde x_{B, m} - \tilde x_{B, n})}}$
%\item [] \hspace{-4.2mm}{\bf if} $u_k^{(\ell)} \!\! < \! \Big(1+ (\tanh J')^{1-2s} (\tanh H')^{2(1- \tilde x_{B, m} - \tilde x_{B, n})}\Big)^{-1}$
%\item [] \hspace{3.0mm}  $\tilde x_{A,k}^{(\ell)} = s$
%\item [] \hspace{3.0mm}  $\tilde x_{B, m} = \tilde x_{B, m} \oplus 1$
%\item [] \hspace{3.0mm}  $\tilde x_{B, n} \, = \tilde x_{B, n} \, \oplus 1$
%%\item [] {\bf else}
%%\item [] \hspace{4.5mm}  $\tilde x_{A, k}^{(\ell)} = s \oplus 1$
%%\item [] \hspace{4.5mm}  {\bf if} $k = |\B|$
%%\item [] \hspace{10mm}  $\tilde x_{B, m}^{(\ell)} = \tilde x_{B, m}^{(\ell -1)} \oplus 1$
%%\item [] \hspace{10mm}  $\tilde x_{B, n}^{(\ell)} \; = \tilde x_{B, n}^{(\ell-1)} \oplus 1$
%%\item [] \hspace{4.5mm}  {\bf else} 
%%\hfill$\blacksquare$
%%\item[] \rule{81.4mm}{1.3pt}
%\end{itemize}
%
%\vspace{3mm}
The number of levels $V$ should be sufficiently large to
ensure that intermediate target distributions are close enough 
and estimating $Z_{\text{d}}(J, |H|^{\alpha_V})$ 
is feasible (see also~\cite[Section 3]{SM:2008}).

%%%%%%%%%%%%%%%%%%%%%%%%%%%%%%%%%%%%%%%%%%%%

%%%%%%%%%%%%%%%%%%%%%%%%%%%%%%%%%%%%%%%%%%%%

%\subsection{3D Ising model} 
%%in an external magnetic field}
%\label{sec:NumIsing3D}
%
%In Section~\ref{sec:NumIsing3D}, we consider 3D Ising models defined
%on a cubic lattice. 
%
%The method can be applied to ferromagnetic 3D Ising models in an
%external field.
%In a model of size $N = 10\times 10\times 10$,
%we set $J_{k,\ell} \overset{\text{i.i.d.}}{\sim} \calU[1.0, 2.0]$ 
%and $H = -1.5$. For one instance of the Ising model, simulation results 
%obtained from importance 
%sampling (solid lines) and uniform sampling (dashed lines) on the dual 
%factor graph are shown in \Fig{fig:FerIsing3D}, where the estimated 
%free energy per site, i.e., $\frac{1}{N}\ln Z$, is about $5.451$.
%
%\begin{figure}[t!]
%\includegraphics[width=1.01\linewidth]{P3D}
%\caption{\label{fig:FerIsing3D}%
%Estimated free energy per site vs.\ the number of samples
%for a $10\times 10\times 10$ ferromagnetic Ising model in an external field with periodic
%boundary conditions, with
%$J\sim\calU[1.0, 2.0]$ and $H = -1.5$. 
%The plot shows five different sample paths obtained from importance 
%sampling (solid lines) and five different sample paths obtained from uniform 
%sampling (dashed lines)
%on the dual factor graph.}
%\end{figure}

\newcommand{\IT}{IEEE Trans.\ Information Theory}
\newcommand{\CASI}{IEEE Trans.\ Circuits \& Systems~I}
\newcommand{\COM}{IEEE Trans.\ Comm.}
\newcommand{\COMLet}{IEEE Commun.\ Lett.}
\newcommand{\COMMag}{IEEE Communications Mag.}
\newcommand{\ETT}{Europ.\ Trans.\ Telecomm.}
\newcommand{\SPMag}{IEEE Signal Proc.\ Mag.}
\newcommand{\ProcIEEE}{Proceedings of the IEEE}

\end{document}